\documentclass[]{llncs}
\usepackage{graphicx}
\usepackage{epsfig}
\usepackage{url}
\usepackage{float}
\usepackage{color}
\usepackage{colortbl}

\newcommand {\secref}[1]   {Section~\ref{#1}}

\newcommand {\figref}[1]   {Figure~\ref{#1}}

\newcommand {\eg}          {e.\,g.\ }
\newcommand {\ie}          {i.\,e.\ }
\newcommand {\cf}          {cf.\ }
\newcommand {\bi}          {\begin{itemize}}
\newcommand {\ei}          {\end{itemize}}

\floatstyle{plain}

\begin{document}

\title{UNICORE - From Project Results\\to Production Grids}

\author{A. Streit, D. Erwin, Th. Lippert, D. Mallmann, R. Menday, \\
M. Rambadt, M. Riedel, M. Romberg, B. Schuller, Ph. Wieder}

\institute{John von Neumann-Institute for Computing (NIC)\\
Forschungszentrum J\"{u}lich (FZJ)\\
52425 J\"{u}lich, Germany\\
E-mail: {\tt \{a.streit,d.erwin,th.lippert,d.mallmann,r.menday,\\
m.riedel,m.rambadt,m.romberg,b.schuller,ph.wieder\}@fz-juelich.de} }

\maketitle

\begin{abstract}
The UNICORE Grid-technology provides a seamless, secure and intuitive access to distributed Grid resources. In
this paper we present the recent evolution from project results to production Grids. At the beginning UNICORE was
developed as a prototype software in two projects funded by the German research ministry (BMBF). Over the
following years, in various European-funded projects, UNICORE evolved to a full-grown and well-tested Grid
middleware system, which today is used in daily production at many supercomputing centers worldwide. Beyond this
production usage, the UNICORE technology serves as a solid basis in many European and International research
projects, which use existing UNICORE components to implement advanced features, high level services, and support
for applications from a growing range of domains. In order to foster these ongoing developments, UNICORE is
available as open source under BSD licence at SourceForge, where new releases are published on a regular basis.
This paper is a review of the UNICORE achievements so far and gives a glimpse on the UNICORE roadmap.
\end{abstract}


\section{Introduction}
\label{sec:intro}
End of 1998 the concept of ``Grid computing" was introduced in the monograph ``The Grid: Blueprint for a New
Computing Infrastructure" by I. Foster and C. Kesselman \cite{Foster:1998:GBN}. Two years earlier, in 1997, the
development of the UNICORE - Uniform Interface to Computing Resources - system was initiated to enable German
supercomputer centers to provide their users with a {\em seamless, secure, and intuitive access} to their
heterogeneous computing resources. Like in the case of the Globus Toolkit\textsuperscript{\textregistered}
\cite{Foster:1997:GMI} UNICORE was started before ``Grid Computing" became the accepted new paradigm for
distributed computing.

The UNICORE vision was proposed to the German Ministry for Education and Research (BMBF) and received funding. A
first prototype was developed in the UNICORE\footnote{funded in part by BMBF grant 01 IR 703, duration: August
1997 - December 1999} project \cite{Erwin:2000:UNI}. The foundations for the current production version were laid
in the follow-up project UNICORE Plus\footnote{funded in part by BMBF grant 01 IR 001 A-D, duration: January 2000
- December 2002} \cite{Erwin:2003:UNI}, which was successfully completed in 2002. Since then UNICORE was used in
operation at German supercomputing centers and became a solid basis for numerous European projects. In this paper
we will describe the evolution of UNICORE from a prototype software developed in research projects to a Grid
middleware used today in the daily operation of production Grids.

Although already set out in the initial UNICORE project proposal in 1997, the goals and objectives of the UNICORE
technology are still valid:
\bi
\item Foremost, the aim of UNICORE is to hide the rough edges resulting from different hardware architectures, vendor
specific operating systems, incompatible batch systems, different application environments, historically grown
computer center practices, naming conventions, file system structures, and security policies -- just to name the
most obvious.
\item Equally, security is a constituent part of UNICORE's design relying on X.509 certificates for the
authentication of users, servers, and software, and the encryption of the communication over the internet.
\item Finally, UNICORE is usable by scientists and engineers without having to study vendor or site-specific
documentation. A Graphical User Interface (GUI) is available to assist the user in creating and managing jobs.
\ei

Additionally, several basic conditions are met by UNICORE: the Grid middleware supports operating systems and
batch systems of all vendors present at the partner sites. In 1997 these were for instance large Cray T3E systems,
NEC and Hitachi vector machines, IBM SP2s, and smaller Linux clusters. Nowadays the spectrum is even broader, of
course with modern hardware, such as IBM p690 systems. The deployed software has to be non-intrusive, so that it
does not require changes in the computing centers hard- and/or software infrastructure. Maintaining site autonomy
is still a major issue in Grid computing, when aspects of acceptability and usability in particular from the
system administrator's point of view are addressed. In addition to UNICORE's own security model, site-specific
security requirements (\eg firewalls) are supported.

Near the end of the initial funding period of the UNICORE Plus project, a working prototype was available, which
showed that the initial concept works. By combining innovative ideas and proven components over the years, this
first prototype evolved to a {\em vertically integrated} Grid middleware solution.

The remainder of this paper is structured as follows. In section 2 the architecture of UNICORE and its core
features are described. European funded projects, which use UNICORE as a basis for their work are described in
Section 3, and in Section 4 the usage of UNICORE in production is described. Section 5 gives an outlook on the
future development of UNICORE. The paper closes with conclusions and acknowledgements.


\section{The Architecture of UNICORE}
\label{sec:arch}
\begin{figure}[htb]
\centering
\includegraphics[width=\textwidth]{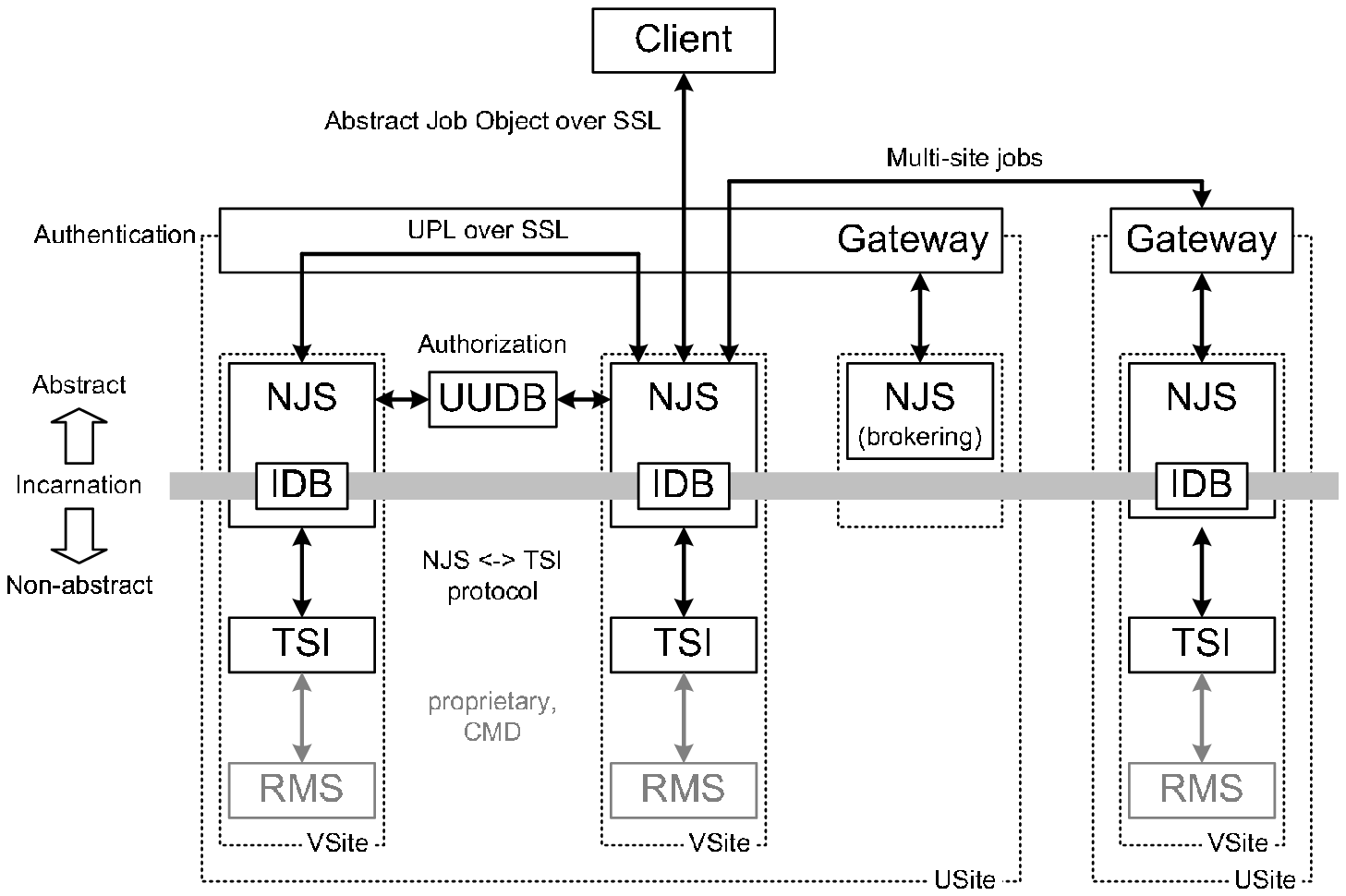}
\caption{The UNICORE architecture.} \label{fig:UNICORE-arch}
\end{figure}

\figref{fig:UNICORE-arch} shows the layered Grid architecture of UNICORE consisting of user, server and target
system tier \cite{Romber:2002:UGI}. The implementation of all components shown is realized in Java. UNICORE meets
the Open Grid Services Architecture (OGSA) \cite{Foster:2003:TPG} concept following the paradigm of 'Everything
being a Service'. Indeed, an analysis has shown that the basic ideas behind UNICORE already realizes this paradigm
\cite{Snelling:2002:UGI,Snelling:2003:UOG}.

\subsection{User Tier}
\label{sec:arch-userT}
The UNICORE Client provides a graphical user interface to exploit the entire set of services offered by the
underlying servers. The client communicates with the server tier by sending and receiving Abstract Job Objects
(AJO) and file data via the UNICORE Protocol Layer (UPL) which is placed on top of the SSL protocol. The AJO is
the realization of UNICORE's job model and central to UNICORE's philosophy of abstraction and seamlessness. It
contains platform and site independent descriptions of computational and data related tasks, resource information
and workflow specifications along with user and security information. AJOs are sent to the UNICORE Gateway in form
of serialized and signed Java objects, followed by an optional stream of bytes if file data is to be transferred.

The UNICORE client assists the user in creating complex, interdependent jobs that can be executed on any UNICORE
site (Usite) without requiring any modifications. A UNICORE job, more precisely a job group, may recursively
contain other job groups and/or tasks and may also contain dependencies between job groups to generate job
workflows. Besides the description of a job as a set of one or more directed a-cyclic graphs, conditional and
repetitive execution of job groups or tasks are also included. For the monitoring of jobs, their status is
available at each level of recursion down to the individual task. Detailed log information is available to analyze
potential error conditions. At the end of the execution of the job it is possible to retrieve the {\sf stdout} and
{\sf stderr} output of the job. Data management functions like import, export, and transfer are available through
the GUI as explicit tasks. This allows the user to specify data transfer from one target system to another (\eg
for workflows), from or to the local workstation before or after the execution of a job, or to store data
permanently in archives.

The previously described features already provide an effective tool to use resources of different computing
centers both for capacity or capability computing, but many scientists and engineers use application packages. For
applications without a graphical user interface, a tool kit simplifies the development of a custom built UNICORE
plug-in. Over the years many plug-ins were developed, so that plug-ins already exist for many standard scientific
applications, as \eg for CPMD (Car-Parrinello Molecular Dynamics) \cite{Huber:2001:SCP}, Fluent or MSC Nastran.

\subsection{Server Tier}
\label{sec:arch-serverT}
The server tier contains the Gateway and the Network Job Supervisor (NJS). The Gateway controls the access to a
Usite and acts as the secure entry point accepting and authenticating UPL requests. A Usite identifies the
participating organization (\eg a supercomputing center) to the Grid with a symbolic name that resolves into the
URL of the Gateway. An organization may be part of multiple Grids offering the same or different resources to
different communities. The Gateway forwards incoming requests to the underlying Network Job Supervisor (NJS) of a
virtual site (Vsite) for further processing. The NJS represents resources with a uniform user mapping scheme and
no boundaries like firewalls between them.

A Vsite identifies a particular set of resources at a Usite and is controlled by a NJS. A Vsite may consist of a
single supercomputer, \eg a IBM p690 System with LoadLeveler, or a Linux cluster with PBS as resource management
system. The flexibility of this concept supports different system architectures and gives the organization full
control over its resources. Note that, there can be more than one Vsite inside each USite as depicted in
\figref{fig:UNICORE-arch}.

The NJS is responsible for the virtualization of the underlying resources by mapping the abstract job on a
specific target system. This process is called ``incarnation" and makes use of the Incarnation Database (IDB).
System-specific data are stored in the IDB describing the software and hardware infrastructure of the system.
Among others, the available resources like software, incarnation of abstract commands (standard UNIX command like
rm, cp, ...) and site-specific administrative information are stored. In addition to the incarnation the NJS
processes workflow descriptions included in an AJO, performs pre- and post-staging of files and authorizes the
user via the UNICORE User Database (UUDB). Typically the Gateway and NJS are running on dedicated secure systems
behind a firewall, although the Gateway could be placed outside a firewall or in a demilitarized zone.

\subsection{Target System Tier}
\label{sec:arch-targetsysT}
The Target System Interface (TSI) implements the interface to the underlying supercomputer with its resource
management system. It is a stateless daemon running on the target system and interfacing with the local resource
manager realized either by a batch system like PBS \cite{openPBS:url} or CCS \cite{Hovestadt:2003:SHR}, a batch
system emulation on top of \eg Linux, or a Grid resource manager like Globus' GRAM
\cite{GRAM:url,Menday:2003:GEU}.

\subsection{Single Sign-On}
\label{sec:arch-signon}
The UNICORE security model relies on the usage of permanent X.509 certificates issued by a trusted Certification
Authority (CA) and SSL based communication across `insecure' networks. Certificates are used to provide a single
sign-on in the client. The client unlocks the user's keystore when it is first started, so that no further
password requests are handed to the user. All authentication and authorization is done on the basis of the user
certificate. At each UNICORE site user certificates are mapped to local accounts (standard UNIX uid/gid), which
may be different at each site, due to existing naming conventions. The sites retain full control over the
acceptance of users based on the identity of the individual -- the distinguished name -- or other information that
might be contained in the certificate. UNICORE can handle multiple user certificates, \ie it permits a client to
be part of multiple, disjoint Grids. It is also possible to specify project accounts in the client allowing users
to select different accounts for different projects on one execution system or to assume different roles with
different privileges.

The private key in the certificate is used to sign each job and all included sub-jobs during the transit from the
client to sites and between sites. This protects against tampering while the job is transmitted over insecure
internet connections and it allows to verify the identity of the owner at the receiving end, without having to
trust the intermediate sites which forwarded the job.


\section{UNICORE Based Projects}
\label{sec:projects}
During the evolutionary development of the UNICORE technology, many European and international projects have
decided to base their Grid software implementations on UNICORE or to extend the growing set of core UNICORE
functions with new features specific to their project focus. The goals and objectives of projects using UNICORE
are not limited to the computer science community alone. Several other scientific domains such as bio-molecular
engineering or computational chemistry are using the UNICORE technology as the basis of their work. In the
following we present short overviews of goals and objectives of UNICORE-based projects and describe additional
functions and services contributed to the UNICORE development.

\subsection{EUROGRID -- Application Testbed for European Grid Computing}
\label{sec:projects-eurogrid}
In the EUROGRID\footnote{funded in part by EC grant IST-1999-20247, duration: November 2000 - January 2004}
project \cite{eurogrid:url} a Grid network of leading European High Performance Supercomputing centers was
established. Based on the UNICORE technology application-specific Grids were integrated, operated and
demonstrated: \bi
\item Bio-Grid for biomolecular science
\item Meteo-Grid for localized weather prediction
\item CAE-Grid for coupling applications
\item HPC-Grid for general HPC end-users
\ei

As part of the project, the UNICORE software was extended by an efficient data transfer mechanism, resource
brokerage mechanisms, tools and services for Application Service Providers (ASP), application coupling methods,
and an interactive access feature \cite{Mallmann:2001:EAT}. Efficient data transfer is a important issue, as Grids
typically rely on public parts of Internet connections. The available limited bandwidth has to be used efficiently
to reduce the transfer time and the integrity of the transferred data has to be maintained, even if the transfer
is interrupted. Depending on the application domain, additional security and confidentiality concerns need to be
considered. This UNICORE high performance data transfer also uses X.509 certificates for authentication and
encryption. To achieve not only a fast and secure transfer of data, but also high-performance capabilities,
network Quality of Service (QoS) aspects, overlapping of streamed data transfers, and packet assembling and
compression techniques are included.

In order to optimize the selection of resources -- either done by the users manually or by a metascheduler
automatically -- resource brokerage mechanisms and detailed resource description abilities are important. Within
the EUROGRID project, mechanisms were added to UNICORE, which allow users to specify their jobs in an abstract way
improving the overall resource selection and accounting. In particular for the benefit of the industrial user
aspects of security, convenience, and cost efficiency were addressed. To this end, the already existing security
concepts of UNICORE were thoroughly evaluated and assessed as being adequate, hence no additional development had
to be done. The task of the developed resource broker is to match the abstract specification of the users jobs and
their requirements with the available resources in the Grid. The resource broker reports the best match back to
the user including an estimate of the costs, which than allows the user to assign the appropriate resources to the
job. For the suppliers of Grid resources (\eg supercomputing centers) the resource broker allows to specify
information about computational resources, architectures, processing power, storage and archiving facilities,
post-processing facilities like visualization equipment, available software packages, and security guarantees. All
this data is enhanced by billing information.

Supercomputing centers converge from pure providers of raw supercomputing power to Application Service Providers
(ASP) running relevant scientific applications. For accounting and billing purposes the ASP needs to know the
exact resources consumed by each customer in each run. For measuring the usage of supercomputers standard
mechanisms provided by the resource management and operating system can be used, but measuring the usage of
licenses requires a sophisticated approach. For some applications, \eg from the Computer Aided Engineering (CAE)
domain, this includes a connection to the applications licence manager. Establishing a link to the above mentioned
resource broker is required to influence their decisions.

For solving complex problems applications from different domains, \eg fluid-structure or
electromagnetism-structure, need to be coupled. This is established by using the EUROGRID resource broker
functionality and combining it with the available Metacomputing functionality developed in the UNICORE Plus
project (\cf \secref{sec:intro}), which allows different schedulers of compute and application resources to
cooperate. Finally, an interactive access to control and steer running application is needed for many scientific
applications. The interactive use includes an interactive shell to actually login to computing resources using the
UNICORE technology and security infrastructure.

EUROGRID used the UNICORE technology to provide the above described services and functionalities by developing new
components. After the project ended, the developed components were revised and useful additions to the core
UNICORE functions are now part of the available UNICORE software.

\subsection{GRIP -- Grid Interoperability Project}
\label{sec:projects-grip}
Grid computing empowers users and organizations to work effectively in an information-rich environment. Different
communities and application domains have developed distinct Grid implementations some based on published open
standards or on domain and community specific features. GRIP\footnote{funded in part by EC grant IST-2001-32257,
duration: January 2002 - February 2004} \cite{GRIP:url} had the objective to demonstrate that the different
approaches of two distinct grids can successfully complement each other and that different implementations can
interoperate. Two prominent Grid systems were selected for this purpose: UNICORE and Globus\texttrademark
\cite{globus:url}, a toolkit developed in the United States. In contrast to UNICORE, Globus provides a set of APIs
and services which requires more in-depth knowledge from the user. Globus is widely used in numerous international
projects and many centers have Globus installed as Grid middleware.

The objectives of GRIP were: \bi
\item Develop software to enable the interoperation of independently developed Grid solutions
\item Build and demonstrate prototype inter-Grid applications
\item Contribute to and influence international Grid standards
\ei

During the runtime of the GRIP project the Open Grid Service Architecture was proposed by the Global Grid Forum
(GGF) \cite{GGF:url}. The arrival of OGSA also was an opportunity to influence the standards directly which were
to be created and to start developments that allow UNICORE to interoperate not only with Globus but with services
on the Grid in general, once the definition of the services and their interfaces became mature. OGSA did not
change the overall objectives of GRIP, however, it influenced directly some of the technical results.

A basic requirement of GRIP was that the Grid interoperability layer should not change the well-known UNICORE user
environment. As developers from both communities cooperated in the GRIP project, this goal was reached with only
little changes of the UNICORE server components and no changes of the Globus Toolkit. This was achieved by the
development of the so called Globus Target System Interface (Globus TSI), which provides UNICORE-access to
computational resources managed by Globus. The Globus TSI was integrated into a heterogeneous UNICORE and Globus
testbed.

To achieve the main objective of GRIP, the interoperability between UNICORE and Globus and initial OGSA services,
the following elements had to be implemented: \bi
\item The interoperability layer between UNICORE and Globus Version 2
\item The interoperability layer between UNICORE and Globus Version 3
\item The Access from UNICORE to simple Web services as a first step towards full integration of Web services
\item The Interoperability of the certificate infrastructures of UNICORE and Globus
\item A resource broker capable of brokering between UNICORE and Globus resources
\item The Ontology of the resource description on an abstract level
\ei

In GRIP, two important application areas were selected to prove that the interoperability layers work as
specified: \bi
\item Bio-molecular applications were instrumented in such a way that they are Grid-aware in any
Grid environment and capable to seamlessly use UNICORE and Globus managed resources. The techniques developed in
GRIP were designed and implemented in a generalized way to ensure that they can be used in other application
domains as well.
\item A meteorological application, the Relocatable Local Model (RLM), was decomposed in such
a way that the components could execute on the most suitable resources in a Grid, independent of the middleware.
\ei

The results of the GRIP project are important for understanding general interoperability processes between Grid
middleware systems. The experience and knowledge of the GRIP partners allowed to work in many relevant areas
within GGF, like security, architecture, protocols, workflow, production management, and applications, and to
influence the work in GGF.

\subsection{OpenMolGRID -- Open Computing Grid for Molecular Science and Engineering}
\label{sec:projects-openmolgrid}
The OpenMolGRID\footnote{funded in part by EC grant IST-2001-37238, duration: September 2002 - February 2005}
project \cite{OpenMolGrid:url} was focused on the development of Grid enabled molecular design and engineering
applications. {\em In silico} testing \cite{Sild:2005:OAW} has become a crucial part in the molecular design
process of new drugs, pesticides, biopolymers, and biomaterials. In a typical design process $O(10^5)$ to
$O(10^6)$ candidate molecules are generated and their feasibility has to be tested. It is not economical to carry
out experimental testing on all possible candidates. Therefore, computational screening methods provide a cheap
and cost effective alternative to reduce the number of candidates. Over the years Quantitative Structure
Activity/Property Relationship (QSAR/QSPR) methods have been shown to be reliable for the prediction of various
physical, chemical, and biological activities \cite{Karelson:2000:MDQ}.

QSPR/QSAR relies on the observation that molecular compounds with similar structure have similar properties. For
each specific application a set of molecules is needed for which the target property is known. This requires
searching globally distributed information resources for appropriate data. For the purpose of exploring molecular
similarity, descriptors are calculated from the molecular structure. Thousands of molecular descriptors have been
proposed and are used to characterize molecular structures with respect to different properties. Their calculation
puts high demands on computer resources and requires high-performance computing.

Based on this complex application the objectives of the OpenMolGRID project were defined as:
\bi
\item Development of tools for secure and seamless access to distributed information and computational methods
relevant to molecular engineering within the UNICORE frame
\item Provision of a realistic testbed and reference application in life science
\item Development of a toxicity prediction model validated with a large experimental set
\item Provision of design principles for next generation molecular engineering systems.
\ei
In particular this included to use UNICORE to automatize, integrate, and speed-up the drug discovery pipeline.

The OpenMolGRID project addressed the objectives above by defining abstraction layers for data sources (databases)
and methods (application software), and integrating all necessary data sources (\eg ECOTOX \cite{ecotox:url}) and
methods (\eg 2D/3D Molecular Structure Conversion and Optimization, Descriptor Calculation, Structure Enumeration)
into UNICORE. The project developed application specific user interfaces (plug-ins) and a mechanism to generate a
complete UNICORE Job from an XML workflow specification. This so called Meta-Plug-in takes care of including all
auxiliary steps like data format transformation and data transfers into the job, distributing data parallel tasks
over available computational resources, and allocating resources to the tasks. Thereby the molecular design
process was significantly improved as the time to build QSAR/QSPR models, the probability for mistakes, and the
variability of results was reduced. In addition a command line client (CLC) for UNICORE was developed to enable
the data warehouse to use Grid resources for its data transformation processes. The CLC offers the generation of
UNICORE jobs from XML workflow description as well as the job submission, output retrieval, status query, and job
abortion. The CLC consists of commands, an API, and a queuing component.

Besides the technical achievements of OpenMolGRID and the added value for pharmaceutical companies its results
will contribute to the standardization of QSAR models.

\subsection{VIOLA -- Vertically Integrated Optical Testbed for Large Applications}
\label{sec:projects-viola}
The aim of the VIOLA\footnote{funded in part by BMBF grant 01AK605F, duration: May 2004 - April 2007} project
\cite{VIOLA:url} is to build up a testbed with the latest optical network technology (multi 10 Gigabit Ethernet
links). The goals and objectives of VIOLA are: \bi
\item Testing of new network components and network architectures
\item Development and testing of software for dynamic bandwidth management
\item Interworking of network technology from different manufacturers
\item Development and testing of new applications from the Grid and Virtual Reality (VR) domain
\ei

The performance of the new network technology is evaluated with different scientific applications that need a very
high network performance and network flexibility. UNICORE is used to build up the Grid on top of the hardware
without taking fundamental software modifications. Only an interface to the meta-computer software library
MetaMPICH \cite{METAMPICH:url} needs to be integrated into UNICORE. Grid applications from the High Performance
Supercomputing and Virtual Reality domain are enhanced for an optimized usage of the available bandwidth and the
provided Quality of Service classes. In this context a Meta-Scheduler framework is developed, which is able to
handle complex workflows and multi-site jobs by coordinating supercomputers and the network connecting them.

\begin{figure}[htb]
\centering
\includegraphics[width=\textwidth]{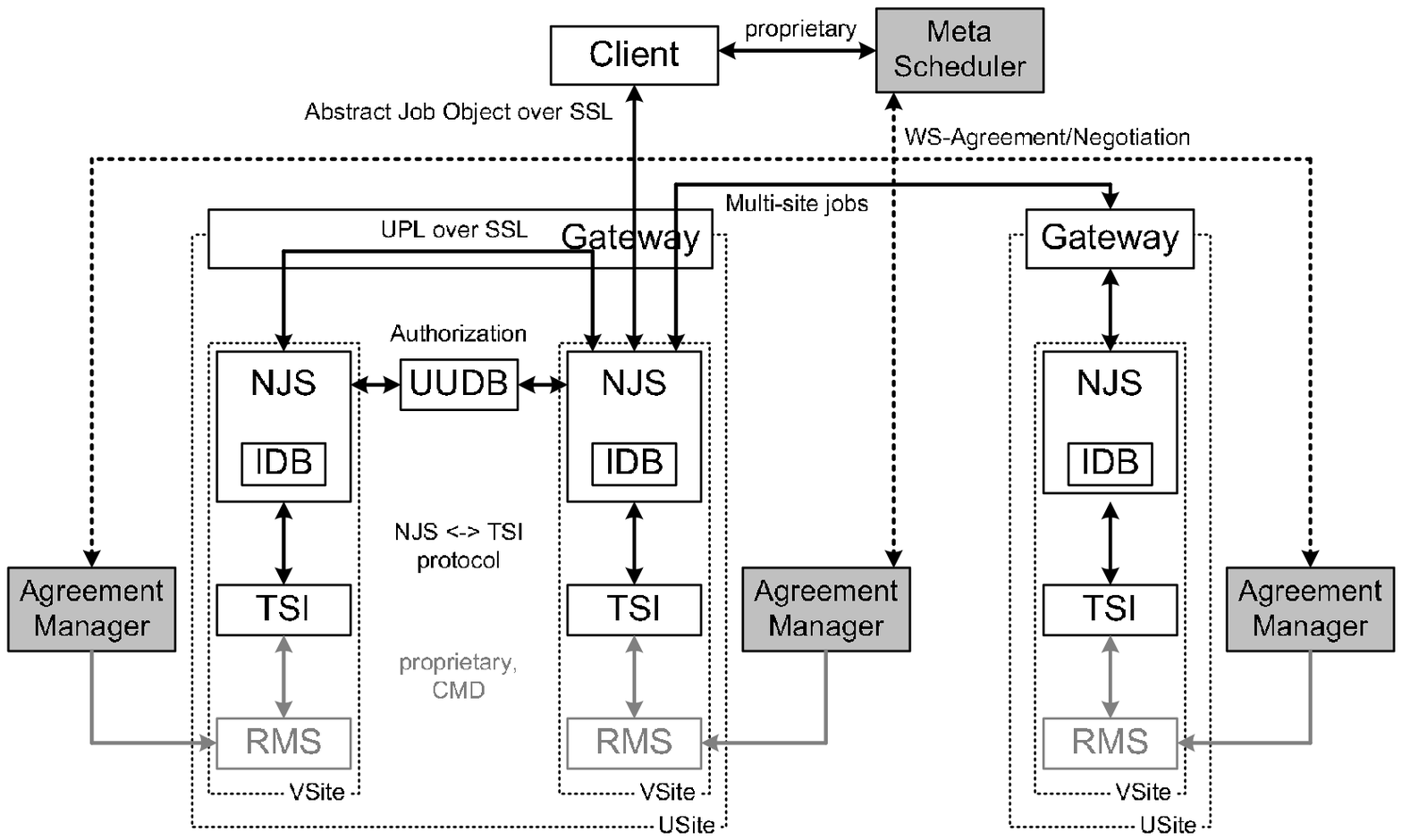}
\caption{The VIOLA Meta-Scheduler architecture.} \label{fig:VIOLA-supersched}
\end{figure}

VIOLA's first generation Meta-Scheduler architecture focuses on the scheduling functionality requiring only
minimal changes to the UNICORE system. As depicted in \figref{fig:VIOLA-supersched}, the system comprises the
Agreement Manager, the Meta-Scheduler itself \cite{Quecke:2000:MAR}, and a Meta-Scheduling plug-in (which is part
of the client and not pictured separately). Before submitting a job to a Usite (\cf \secref{sec:arch-serverT}),
the Meta-Scheduling plug-in and the Meta-Scheduler exchange the data necessary to schedule the resources needed.
The Meta-Scheduler is then (acting as an Agreement Consumer in WS--Agreement terms \cite{GRAAP:url}) contacting
the Agreement Manager to request a certain level of service, a request which is translated by the Manager into the
appropriate resource management system commands. In case of VIOLA's computing resources the targeted resource
management system is the EASY scheduler. Once all resources are reserved at the requested time the Meta-Scheduler
notifies the UNICORE Client via the Meta-Scheduling plug-in to submit the job. This framework will also be used to
schedule the interconnecting network, but potentially any resource can be scheduled if a respective Agreement
Manager is implemented and the Meta-Scheduling plug-in generates the necessary scheduling information. The
follow-on generation of the Meta-Scheduling framework will then be tightly integrated within UNICORE/GS (\cf
\secref{sec:future-unigrids-GS}).

\subsection{NaReGI -- National Research Grid Initiative}
\label{sec:projects-naregi}
The Japanese NaReGI project \cite{NAREGI:url} includes the UNICORE technology as the basic middleware for research
and development. NaReGI is a collaboration project between industry, academia, and government. The goals and
objectives are: \bi
\item Establishment of a national Japanese research Grid infrastructure
\item Revitalization of the IT industry through  commercialization of Grid middleware and strengthened
international competitiveness
\item Dissemination of Grid environments throughout industry
\item Trailblazing the standardization of Grid technology
\item Cultivation of human resources specializing in IT technology for Grids
\ei

Similar to the GRIP project (\cf \secref{sec:projects}) where an interoperability layer between UNICORE and Globus
Toolkit 2 and 3 was developed, the NaReGI project plans to implement such a layer between UNICORE and Condor
\cite{CONDOR:url}, called UNICONDORE. This interoperability layer will allow to submit jobs from the UNICORE
client to Condor pools and to use Condor commands to submit jobs to UNICORE managed resources.

In the first phase of the NaReGI testbed UNICORE provides access to about 3000 CPUs in total with approximately 17
TFlops of peak performance. It is expected to increase the integrated peak performance to 100+ TFlops by the end
of the project in 2007.


\section{UNICORE in Production}
\label{sec:production}
From its birth in two German BMBF-funded projects to its extensive use and further development in a variety of EU
and BMBF research projects, the UNICORE technology ran through an evolutionary process transforming from an
initial prototype software to a powerful production Grid middleware.

\subsection{UNICORE@SourceForge}
\label{sec:production-SF}
Since May 2004, the UNICORE technology with all its components is available as open source software under the BSD
license. It can be downloaded from the SourceForge repository. Besides the core developers of UNICORE (namely
Fujitsu Laboratories of Europe, Intel Germany and the Research Center J\"{u}lich), there are numerous contributors
from all over the world, \eg Norway, Poland, China and Russia. The Web site \cite{UNICORE-sourceforge:url} offers
a convenient entry point for interested users and developers. In the download section the UNICORE software is
bundled in different packages, \eg the client package and individual packages for the different server components
Gateway, NJS, TSI/IDB, UUDB (\cf \secref{sec:arch}), and plug-ins. Until January 2005 more than 2800 downloads of
UNICORE are counted.

A tracker section linked on the Web site establishes a communication link to the core developer community. The
corresponding mailing lists allow users to report bugs, to request new features, and to get informed about bug
fixes or patches. For the announcement of new software releases a separate mailing list was created. The Grid team
at the Research Center J\"{u}lich is responsible for UNICORE@SourceForge. Its work includes coordinating and driving
the development effort, and producing consolidated, stable, and tested releases of the UNICORE software.

\subsection{Production System on Jump}
\label{sec:production-Jump}
Since July 2004 UNICORE is established as production software to access the supercomputer resources of the John
von Neumann-Institute for Computing (NIC) at the Research Center J\"{u}lich. These are the 1312-processor IBM p690
cluster (Jump) \cite{Jump:url}, the Cray SV1 vector machine, and a new Cray XD1 cluster system. As an alternative
to the standard SSH login, UNICORE provides an intuitive and easy way for submitting batch jobs to the systems.
The academic and industrial users come from all over Germany and from parts of Europe. The applications come from
a broad field of domains, \eg astrophysics, quantumphysics, medicine, biology, chemistry, and climate research,
just to name the largest user communities. A dedicated, pre-configured UNICORE client with all required
certificates and accessible Vsites is available for download. This alleviates the installation and configuration
process significantly. Furthermore, an online installation guide including a certificate assistant, an user
manual, and example jobs help users getting started.

To provide the NIC-users with adequate certificates and to ease the process of requesting and receiving a
certificate, a certificate authority (CA) was established. User certificate requests are generated in the client
and have to be send to the CA. Since introduction of UNICORE at NIC, more than 120 active users requested a
UNICORE user certificate.

A mailing list serves as a direct link of the users to UNICORE developers in the Research Center J\"{u}lich. The list
allows to post problems, bug reports, and feature requests. This input is helpful in enhancing UNICORE with new
features and services, in solving problems, identifying and correcting bugs, and influences new releases
of UNICORE available at SourceForge.

\subsection{DEISA -- Distributed European Infrastructure for Scientific Applications}
\label{sec:production-DEISA}
Traditionally, the provision of high performance computing resources to researchers has traditionally been the
objective and mission of national HPC centers.On the one hand, there is an increasing global competition between
Europe, USA, and Japan with growing demands for compute resources at the highest performance level, and on the
other hand stagnant or even shrinking budgets. To stay competitive major investments are needed every two years --
an innovation cycle that even the most prosperous countries have difficulties to fund.

To advance science in Europe, eight leading European HPC centers devised an innovative strategy to build a
Distributed European Infrastructure for Scientific Applications (DEISA) \cite{DEISA:url}. The centers join in
building and operating a tera-scale supercomputing facility. This becomes possible through deep integration of
existing national high-end platforms, tightly coupled by a dedicated network and supported by innovative system
and grid software. The resulting virtual distributed supercomputer has the capability for natural growth in all
dimensions without singular procurements at the European level. Advances in network technology and the resulting
increase in bandwidth and lower latency virtually shrink the distance between the nodes in the distributed
super-cluster. Furthermore, DEISA can expand horizontally by adding new systems, new architectures, and new
partners thus increasing the capabilities and attractiveness of the infrastructure in a non-disruptive way.

By using the UNICORE technology, the four core partners of the projects have coupled their systems using virtually
dedicated 1 Gbit/s connections. The DEISA super-cluster currently consists of over 4000 IBM Power 4 processors and
416 SGI processors with an aggregated peak performance of about 22 teraflops. UNICORE provides the seamless,
secure and intuitive access to the super-cluster.

The Research Center J\"{u}lich is one of the DEISA core partners and is responsible for introducing UNICORE as Grid
middleware at all partner sites and for providing support to local UNICORE administrators.

\begin{figure}[htb]
\centering
\includegraphics[width=0.95\textwidth]{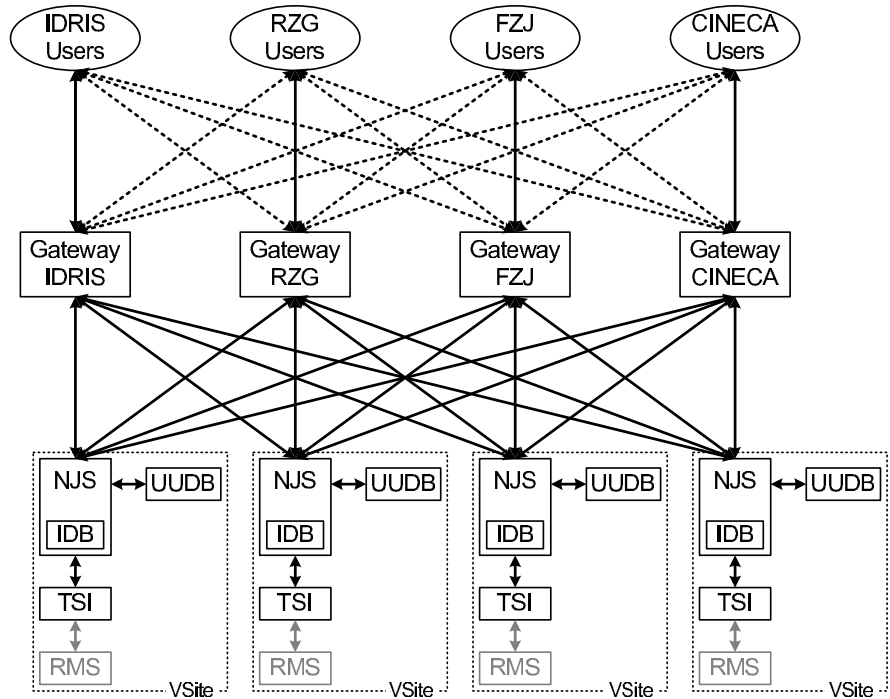}
\caption{The DEISA architecture.} \label{fig:DEISA-architecture}
\end{figure}

All DEISA partners have installed the UNICORE server components Gateway, NJS, TSI, and UUDB to access the local
supercomputer resources of each site via UNICORE. \figref{fig:DEISA-architecture} shows the DEISA UNICORE
configuration. For clarity only four sites are shown. At each site, a Gateway exists as an access to the DEISA
infrastructure. The NJSs are not only registered to their local Gateway, but to all other Gateways at the partner
sites as well. Local security measures like firewall configurations need to consider this, by permitting access to
all DEISA users and NJSs. This fully connected architecture has several advantages. If one Gateway has a high
load, access to the high performance supercomputers through DEISA is not limited. Due to the fully connected
architecture, no single point of failure exists and the flexibility is increased.

The DEISA partners operate different supercomputer architectures, which are all accessible through UNICORE.
Initially all partners with IBM p690 clusters are connected to one large virtual supercomputer. In a second step
other supercomputers of different variety are connected to DEISA, making the virtual supercomputer heterogeneous.
UNICORE can handle this, as it is designed to serve such heterogeneous architectures in a seamless, secure, and
intuitive way.

In December 2004 a first successful UNICORE demonstration between the four DEISA core sites FZJ (Research Center
J\"{u}lich, Germany), RZG (Computing Center Garching, Germany), CINECA (Italian Interuniversity Consortium, Italy) and
IDRIS (Institute for Development and Resources in Intensive Scientific Computing, France) was given. Different
parts of a distributed astrophysical application were generated and submitted with UNICORE to all four sites.

The experience and knowledge of the researchers, developers, users, and administrators in working with UNICORE in
the DEISA project on a large production platform will be used as useful input for future developments of the
UNICORE technology. A close synchronization with the UniGrids project (\cf \secref{sec:future-unigrids}) is
foreseen.


\section{Future of UNICORE}
\label{sec:future}
The current UNICORE software implements a vertically integrated Grid architecture providing seamless access to
various resources. Every resource is statically integrated into the UNICORE Grid by providing an interface to the
appropriate resource manager.

One of the benefits Web services will bring to Grid computing is the concept of loosely coupled distributed
services. Merging the idea of ``everything being a service'' with the achievements of the Grid community led to
Grid services, enabling a new approach to the design of Grid architectures. The adoption of XML and the drive for
standardization of the Open Grid Service Architecture provide the tools to move closer to the promise of
interoperable Grids. A demonstrator validated the correspondence of UNICORE's architectural model with the
OGSA/OGSI (Open Grid Service Infrastructure \cite{Tuecke:2003:OGSI}) approach, which encouraged the development of
an OGSA/OGSI compliant UNICORE Grid architecture in the GRIP project (\cf \secref{sec:projects-grip}).

In \cite{Menday:2003:GEU} UNICORE is examined for the evolution of a Grid system towards a service oriented Grid,
primarily focussing on architectural concepts and models. Based on the current architecture and the enhancements
provided by GRIP, first steps already integrate Web services into UNICORE. This included the provision of OGSI
compliant port types parallel to the proprietary ones as well as the design of XML based protocols. This work was
continued in the UniGrids project.

As mentioned above the development of a Grid middleware is an continuous process of integrating new features,
services, and adapting to emerging standards, and UNICORE is no exception. In the following we present new
developments, some technical details, and report on projects, which enhance the UNICORE technology to serve the
demands of the Grid in the future \cite{Jeffrey:2004:NGG}.

\subsection{UniGrids -- Uniform Interface to Grid Services}
\label{sec:future-unigrids}
The strength of the UNICORE architecture is well-proven as described above. The rapid definition and adoption of
OGSA allow the UNICORE development community to re-cast and extend the concepts of UNICORE through the use of Web
services technologies. The goal of the UniGrids\footnote{funded in part by EC grant IST-2002-004279, duration:
July 2004 - June 2006} project \cite{UniGrids:url} is to lift UNICORE on an architecture of loosely-coupled
components while keeping its 'end-to-end' nature.

Thus, the integration of Web services techniques and UNICORE, which already started in the GRIP project (\cf
\secref{sec:projects-grip}), will continue in the \mbox{UniGrids} project. Interoperability, through adopting and
influencing standards, form the philosophical foundation for UniGrids. The project aims to transform UNICORE into
a system with interfaces that are compliant with the Web Services Resource Framework (WS-RF) \cite{WSRF:url} and
that interoperate with other WS-RF compliant software components.

Such an approach offers great advantages both for the ease of development of new components by aggregation of
services and through the integration of non-UNICORE components into the standards-based infrastructure.

In this sense, work is continuing in the following areas: \bi
\item Development of a compliant WS-RF hosting environment used for publishing UNICORE job and file services as
Web services.
\item Support of dynamic virtual organizations by enhancing the UNICORE security infrastructure to allow different
usage models such as delegation and collective authorization.
\item Development of translation mechanisms, such as resource ontologies, to interoperate with other OGSA compliant
systems. Support for Grid economics by developing a Service Level Agreement (SLA) framework and cross-Grid
brokering services.
\item Development and integration of generic software components for visualization and steering of simulations (VISIT
\cite{visit:url}), device monitoring and control, and tools for accessing distributed data and databases.
\ei

Applications from the scientific and industrial domain, like biomolecular and computational biology, geophysical
depth imaging by oil companies, automotive, risk-management, energy, and aerospace are used to prove the
developments in UniGrids.

The development in the UniGrids project will lead to UNICORE/GS, which follows the architecture of OGSA through
the standardization of WS-RF and related work like \eg the Web Services Notification technology \cite{WSN:url}.
The results will be made available under an open source BSD license.

\subsubsection{UNICORE/GS}
\label{sec:future-unigrids-GS}
Web service technology, and in particular the WS-RF, forms the basis for the UNICORE/GS software. WS-RF is the
follow-on to OGSI, but more in line with mainstream Web services architecture \cite{WSARCH:url}. Based on this new
technology, UNICORE/GS will retain its key characteristics of seamlessness, security, and intuitiveness from both
the user and administrative perspective, but will be built on a service oriented framework. This means that there
is a loosening of the coupling between the components of the system. UNICORE/GS keeps the classical UNICORE
topology of Usites, each containing a number of Vsites, but provides a new framework for integrating other
services and providing common infrastructure functionality as services. This has the implication that new services
will be easily integrated into the UNICORE/GS environment. Conversely, UNICORE/GS will be well-prepared to make
use of external services.

The WS-RF technology is used to model core functionalities such as job submissions and file transfers as
WS--Resources. These services are accessible via web service interfaces and thus establishing the UniGrids atomic
services layer. This layer will be realized making extensive use of existing UNICORE server components.

All services in a Usite are accessible through the UniGrids Gateway that provides a secure entrance into the
UNICORE/GS infrastructure. The principal is exactly the same as for classic UNICORE, however, the Gateway now
routes messages according to Web Services Addressing (WS--Addressing) \cite{WSA:url}. Authentication is based on
transport level HTTPS security, although the intention is to move to Web Services Security (WS--Security)
\cite{WSS:url}. Regarding authorized access to resources, the UNICORE User Database (UUDB) will be available as a
service to other services in the Usite, and will form the basis for future work concerning virtual organizations
and fine-grained authorization schemes.

The underlying UniGrids atomic services layer will provide an excellent framework to deploy higher-level services
such as co-allocation schedulers, workflow engines, and services for provision and easy access to data-intensive,
remotely-steerable simulations.

\subsection{NextGrid -- Architecture for Next Generation Grids}
\label{sec:future-nextgrids}
In comparison to the UniGrids project which evolves the existing UNICORE Grid system to a service-oriented one,
the NextGRID\footnote{funded in part by EC grant IST-2002-511563, duration: September 2004 - August 2007}
\cite{nextgrid:url} project aims for the future: The goal is to provide the foundations for the next generation of
Grids. NextGRID is not a project based on the UNICORE architecture or Grid system as-is, but institutions and
people involved in the UNICORE development from the beginning on contribute expertise and experience to NextGRID.

Since it is obvious that there is no such thing as the one and only next generation Grid, and experts envisage the
co-existence of multiple Grids with well-defined boundaries and access points, NextGRID is going to define a Grid
architecture which can be seen as building blocks for Grids. It does not only provide interoperability by-design
between entities which exist within one instantiation of such an architecture, but it also facilitates the
interoperability between different Grids developed according to the NextGRID architecture.

Although developing a Grid one generation ahead, NextGRID is not starting from scratch. Properties to incarnate
and functions to realize future Grids are expertly described in \cite{Priol:2003:NGG} and \cite{Jeffrey:2004:NGG}.
These reports frame NextGRID's architectural development while the Open Grid Services Architecture is going to
define Grid services and their interactions and does therefore make up a staring point for the conceptualization
and design of NextGRID. In addition, regarding the underlying technology and architectural model, NextGRID
propagates the usage of Web Services and the adoption of Service-Oriented Achitecture (SOA) \cite{Erl:2004:SOA}
concepts and models.

NextGRID focuses on security, economic sustainability, privacy/legacy, scalability and usability. The following
properties have the highest priorities when carrying out the following work: \bi
\item Developing an architecture for next generation Grids
\item Implementing and testing prototypes aligned with the concepts and design of the NextGRID architecture
\item Creating reference applications which make use of the NextGRID prototypes
\item Facilitating the transition from scientific- to business-oriented Grids by integrating the means to negotiate
a certain Quality of Service (QoS) level
\item Specifying the methods, processes, and services necessary to dynamically operate Grids across multiple
organizations which comprise heterogeneous resources
\ei

Since the ongoing UNICORE development in projects like UniGrids shares resources as well as the technological
foundation with NextGRID there is a high chance that the outcome of NextGRID will also represent the next step of
UNICORE's evolution.


\section{Conclusion}
\label{sec:conclusion}
In this paper we presented the evolution of the UNICORE technology from a Grid software with prototype character
developed in two German projects to a full-grown, well-tested, widely used and accepted Grid middleware. UNICORE
-- Uniform Interface to Computing Resources -- provides a {\em seamless, secure and intuitive} access to
distributed Grid resources. Although the UNICORE vision was already coined in 1997, the then stated goals and
objectives of hiding the seams of resource usage, incorporating a strong security model, and providing an easy to
use graphical user interface for scientists and engineers are still valid today: to achieve these goals and
objectives, UNICORE is designed as a vertically integrated Grid middleware providing components at all layers of a
Grid infrastructure, from a graphical user interface down to the interfaces to target machines.

Initially developed in the German projects UNICORE and UNICORE Plus, UNICORE was soon established as a promising
Grid middleware in several European projects. In the GRIP project an interoperability layer between UNICORE and
the Globus Toolkit 2 and 3 was developed to demonstrate the interoperability of independently developed Grid
solutions, allowing to build and to demonstrate inter-Grid applications from the bio-molecular and meteorological
domain. In the EUROGRID project, European high performance supercomputing centers joined to extend UNICORE with an
efficient data transfer, resource brokerage mechanisms, ASP services, application coupling methods, and an
interactive access. In addition, a Bio-Grid, Meteo-Grid, CAE-Grid, and HPC-Grid were established to integrate a
variety of application domains. The main objective of the OpenMolGRID project is to provide a unified and
extensible information-rich environment based on UNICORE for solving problems from molecular science and
engineering. In the VIOLA project a vertically integrated testbed with the latest optical network technology is
built up. UNICORE is used as the Grid middleware for enabling the development and testing of new applications in
the optical networked testbed, which provides advanced bandwidth management and QoS features.

With these developments UNICORE grew to a software system usable in production Grids. In this context UNICORE is
deployed in the large German supercomputing centers to provide access to their resources. At the John von
Neumann-Institute for Computing, Research Center J\"{u}lich, many users submit their batch jobs through UNICORE to the
1312-processor 8.9 TFlop/s IBM p690 cluster and the Cray SV1 vector machine. Leading European HPC centers joined
in the project DEISA to build a distributed European infrastructure for scientific applications based on UNICORE
to build and operate a distributed multi tera-scale supercomputing facility.

The future of UNICORE is promising and follows the trend of ``Everything being a Service" by adapting to Open Grid
Service Architecture (OGSA) standards. In this context, the UniGrids project continues the effort of the GRIP
project in integrating the Web Services and UNICORE technology to enhance UNICORE to an architecture of
loosely-coupled components while keeping its ``end-to-end" nature. To this end UNICORE/GS will be developed, which
makes UNICORE compliant with the Web Services Resource Framework (WS-RF).

Today the UNICORE software is available as open source under a BSD licence from SourceForge for download. This
enables the community of core UNICORE developers to grow and makes future development efforts open to the public.


\section{Acknowledgments}
\label{sec:ack}
The work summarized in this paper was done by many people. We gratefully thank them for their past, present, and
future contributions in developing the UNICORE technology. Most of the work described here was supported and
funded by BMBF and different programmes of the European Commission under the respective contract numbers mentioned
above.


\newcommand{\noopsort}[1]{}

\end{document}